# A custom-tailored multi-TW optical electric field for gigawatt soft-x-ray isolated attosecond pulses

High-power optical waveform synthesizer

**Authors**

Bing Xue[1†], Yuuki Tamaru[1,2], Yuxi Fu[1††], Hua Yuan[3], Pengfei Lan[3], Oliver D. Mücke[4], Akira Suda[2], Katsumi Midorikawa[1], and Eiji J. Takahashi[1*]

**Affiliations**

1 RIKEN Center for Advanced Photonics, RIKEN, 2-1 Hirosawa, Wako, Saitama 351-0198, Japan

2 Tokyo University of Science, 2641 Yamazaki, Noda-shi, Chiba-ken 278-8510, Japan

3 School of Physics and Wuhan National Laboratory of Optoelectronics, Huazhong University of Science and Technology, Wuhan, Hubei 430074, China

4 Center for Free-Electron Laser Science, DESY, Notkestraße 85, 22607 Hamburg, Germany

†bing.xue@riken.jp

†† Present address: Xi'an Institute of Optics and Precision Mechanics, Chinese Academy of Sciences, Xi'an, Shaanxi 710119, P.R. China.

*Corresponding author: ejtak@riken.jp

**Abstract**

Since the first isolated attosecond pulse was demonstrated through high-order harmonics generation (HHG) in 2001, researchers' interest in the ultrashort time region has expanded. For example, directly observing the fast motion of an electron in atoms might be possible for the first time using an attosecond pulse. However, one realizes a limitation for related research such as attosecond spectroscopy. The bottleneck is concluded to be the lack of a high-peak-power isolated attosecond pulse source. Therefore, currently, generating an intense attosecond pulse would be one of the highest priority goals. In this paper, we review a TW-class parallel three-channel waveform synthesizer for generating a gigawatt-scale soft-x-ray isolated attosecond pulse (IAP) using HHG. Simultaneously, using several stabilization methods, namely, the low-repetition-rate laser carrier-envelope phase stabilization, Mach–Zehnder interferometer, balanced optical cross-correlator, and beam-pointing stabilizer, we demonstrate a stable 50-mJ three-channel optical-waveform synthesizer with a peak power at the multi-TW level. This optical-waveform synthesizer is capable of creating a stable intense optical field for generating an intense continuum harmonic beam thanks to the successful stabilization of all the parameters. Furthermore, the precision control of shot-to-shot reproducible synthesized waveforms is achieved. Through the HHG process employing a loose-focusing geometry, an intense shot-to-shot stable supercontinuum (50–70 eV) is generated in an argon gas cell. This continuum spectrum supports an IAP with a transform-limited duration of 170 as and a submicrojoule pulse energy, which allows the generation of a GW-scale IAP. Another supercontinuum in the soft-x-ray region with higher photon energy of approximately 100–130 eV is also generated in neon gas from the synthesizer. The transform-limited pulse duration is 106 as. According to this work, the enhancement of HHG output through optimized waveform synthesis is experimentally proved. The high-energy multicycle pulse with 10-Hz repetition rate is proved to have the same controllability for optimized waveform synthesis for HHG as few- or subcycle pulses from a 1-kHz laser.

**MAIN TEXT**

1. **Introduction**

Since the first isolated attosecond pulse (IAP) was demonstrated by high-order harmonic generation (HHG) in 2001 [1], attosecond science has emerged as a vital frontier research area of ultrafast science [2-4]. Theoretically, for generating an IAP in the time domain, the HHG spectrum contains a continuum spectrum, which requires controlling the pulse width of the driving laser within a few optical cycles. However, in actual experiments, driving HHG directly with sufficiently intense laser pulses with a few cycles or even less than a single cycle duration is not easy. Therefore, researchers have proposed and adopted gating methods to filter a relatively long driver pulse (a few cycles) to obtain IAPs [5-11], for example, amplitude gating [1,5], polarization gating [6,7], double optical gating [8,9], and spatial gating [10,11]. Currently, the continuous optimization of existing gating schemes and inventing new gating methods are still major topics in the research on IAP generation [11-13].

The emergence of IAPs provides pivotal contributions for the study of electronic dynamics in complex systems. However, the lack of high-photon-flux and high-peak-power IAPs is considered a restriction for a variety of applications [2-4,14-16] such as IAP pump–IAP probe experiments, subnanometer imaging by IAP, and nonlinear IAP experiments. The reason for the low photon flux is the low energy-conversion efficiency through the HHG process [14]. Currently, most attosecond pump–probe experiments (including streaking measurements) have to employ a two-color scheme to compensate for the energy lack: an IAP plus a near-infrared (NIR) femtosecond pulse [15,16], where the NIR serves as the pump or probe pulse. However, with this scheme, the large ponderomotive energy associated with NIR pulses often causes difficulty in the interpretation of experimental observations. Therefore, a real attosecond pump–attosecond probe experiment remains a dream.

Recently, another route for obtaining a suboptical cycle pulse is proposed by synthesizing multiple pulses with different wavelengths delivered from the same laser system. This scheme is called an optical-waveform synthesizer. Using optical parametric (chirped-pulse) amplification (OP(CP)A) [17-20], multioctave spectral spanning with subcycle-timing jitter control [17] has been experimentally achieved using a synthesizer. Usually, waveform synthesizers can be categorized into two different schemes, namely, sequential and parallel schemes. In the sequential scheme, as used in the parametric pulse synthesizer [19], the full bandwidth is never spatially separated/recombined. Instead, the different copropagating spectral regions are amplified in different amplification stages employing different phase-matching conditions. The benefit is that all wavelengths travel the same path, thus no active stabilization of the relative timing and relative phase is required. However, this benefit is at the same time also a severe drawback if one wants to tailor waveforms from different regions having a relative delay or relative phase, which is more challenging to implement compared with the parallel scheme. The biggest drawback and unsolved challenge of this approach is that the whole bandwidth needs to be handled all the time, while it is limited by the bandwidth of precision chirp-control devices, such as an acousto-optic programmable dispersion filter (Dazzler), which imposes severe limitations on the total bandwidth. On the other hand, precision chirp compensation of parallel synthesizers has been demonstrated for over two octaves in bandwidth [17,20]. In the parallel scheme, different wavelength portions are split into parallel channels. Splitting up the whole bandwidth into narrower-bandwidth channels has the benefit that one can handle narrower bandwidths separately and more flexibly (e.g., in parametric amplification stages, for precision chirp compensation, changing the relative time and relative phase of individual channels), but it requires active stabilization of the synthesis parameters when recombining the channels again finally to create a shot-to-shot stable superposed field $E_{synth}(t)$. The unstable elements are mainly due to mechanical vibrations causing delay jitter of the optical paths between each incident pulse, and phase jitter including carrier-envelope phase (CEP) changes. To achieve sufficiently high active feedback bandwidth for each jitter, a high-repetition laser is usually demanded. By considering the above questions, the main challenge is how to obtain a shot-to-shot stable-waveform electric field through a synthesizer system driving the HHG process.

Chipperfield *et al.* theoretically proposed a "perfect waveform" by synthesized subcycle driver pulses for optimizing the HHG cutoff energy [21] and IAP production [22,23]. Through waveform synthesis, enhancement of the HHG has been studied experimentally and theoretically by other researchers [24-26]. However, the output energies of these synthesizers are limited to the few hundred microjoules class with a kilohertz repetition rate by the fundamental laser source. To generate an intense IAP pulse, synthesizing the well-developed few- or multicycle laser pulses is expected to enable scaling up of the energy and efficiency [13,23]. When we are targeting a microjoule-level IAP generation, which is the basic requirement for real attosecond pump–probe experiments, considering the conversion efficiency for a typical HHG process, a multi-TW-class waveform synthesizer is needed. This requirement leads to the use of high-energy femtosecond lasers with a limited repetition rate (e.g., 10 Hz) for driving the waveform synthesizer, for the currently well-developed laser systems. Moreover, not only power but also stability during long-term data acquisition are required. The technical challenges of developing a stable optical-waveform synthesizer based on such a high-power laser system with a low repetition rate are tremendous. Better performance of the above-mentioned stabilization methods for stabilizing CEP, reducing delay jitter, and beam pointing, demands higher feedback frequency, which is usually a weak point of a high-power laser system with a low repetition rate.

In this review article, we report a multi-TW 10-Hz three-channel parallel parametric waveform synthesizer with complete stabilization of all synthesis parameters for the first time, consisting of pulses with 800-, 1350-, and 2050-nm wavelengths. Using the intense-synthesizer output waveform as a driving source, HHG is generated in a cell filled with argon and neon. Supercontinua are generated in the soft-x-ray region with the cutoff regions in 50–70 eV (argon) and 100–130 eV (neon), respectively. Both of the continua support sub-200 as IAP generation. In particular, the pulse energy for the region of 50–70 eV is over 240 nJ, which indicates GW-scale IAP generation through a synthesizer for the first time. Furthermore, the stability of the synthesized waveform by multicycle pulse synthesis is discussed with the experimental facts and theoretical perspective.

2. **Methods**

The intense synthesizer system built in this work consists of three laser pulses from an optical parametric amplifier (OPA): pump (44 mJ, 800 nm, 30 fs full-width at half-maximum), signal (6 mJ, 1350 nm, 44 fs), and idler (3 mJ, 2050 nm, 88 fs); detailed information on the laser pulses is shown in *Figure S1*. The pulse durations of the component pulses (pump, signal, and idler pulses) of the synthesizer are measured separately using different methods, spectral phase interferometry for direct electric-field reconstruction for the pump pulse [27] and frequency-resolved optical gating for the signal and idler pulses [28]. The pump pulse is from a portion of the energy of the laser source, and the signal and idler pulses are two output pulses from a three-stage infrared OPA, as shown in the system scheme in *Figure 1*. Because the external IR seed pulses for the three-stage OPA are produced by white-light continuum generation, the CEP of the signal pulse is determined by the 800-nm pulse and preserved in the amplified signal pulse. After the delay and polarization control, the three pulses are synthesized by spatial recombination on dichroic mirrors. Therefore, the characteristics of the synthesizer system in this work are multicycle pulse synthesis, with total output power attaining multi-TW level.

The output electric field ($E_{3c}$) of our three-channel parallel synthesizer can be described by the following *Equation (1)*. It contains three terms, which are the coherently added electric fields of each channel (component pulse):

$$E_{3c} = \sum_{m=p,s,i} E_m \exp\left[-2\ln 2 \left(\frac{t+\delta t_m}{\tau_m}\right)^2\right] \cos[\omega_m(t+\delta t_m) + \varphi_m]. \qquad (1)$$

Here, $E_m$ is the electric-field amplitude of the component pulses, $\tau_m$ is their pulse durations, $\omega_m$ is their center frequencies, $\delta t_m$ is the delays of the pulses, and $\varphi_m$ is the initial phases of the pulses. The subscripts *p*, *s*, and *i* correspond to the pump, signal, and idler pulses, respectively. To obtain a shot-to-shot stable optical electrical field experimentally, we need to stabilize all the unstable parameters in *Equation (1)*. When the conditions of the pump laser and OPAs are fixed ($E_m$, $\tau_m$, $\omega_m$) by the laser setup, the remaining changeable components in the equation are the delays ($\delta t_m$) and phases ($\varphi_m$) of all channels only. In addition, the phase of a laser field, called CEP, randomly changes from shot to shot and thus needs to be stabilized. Because three channels are introduced in this work, the delay jitters and phase jitters among each channel also need to be stabilized. By analyzing, we found that there are three main jitter sources, namely, laser CEP, delay jitters due to mechanical vibrations in the optical paths, and the laser energy fluctuations induced delay changes during the white-light generation in the OPA [29]. After introducing the stabilization for laser CEP [30], Mach–Zehnder (MZ) interferometers [31], and the balanced optical cross-correlator (BOC) [17], we directly targeted the delay jitter sources and successfully stabilized the jitters between all three channels in the synthesizer. Details of the stabilization methods are discussed below. The stabilized parameters are listed in **Table 1** for the full stabilization of the synthesizer.

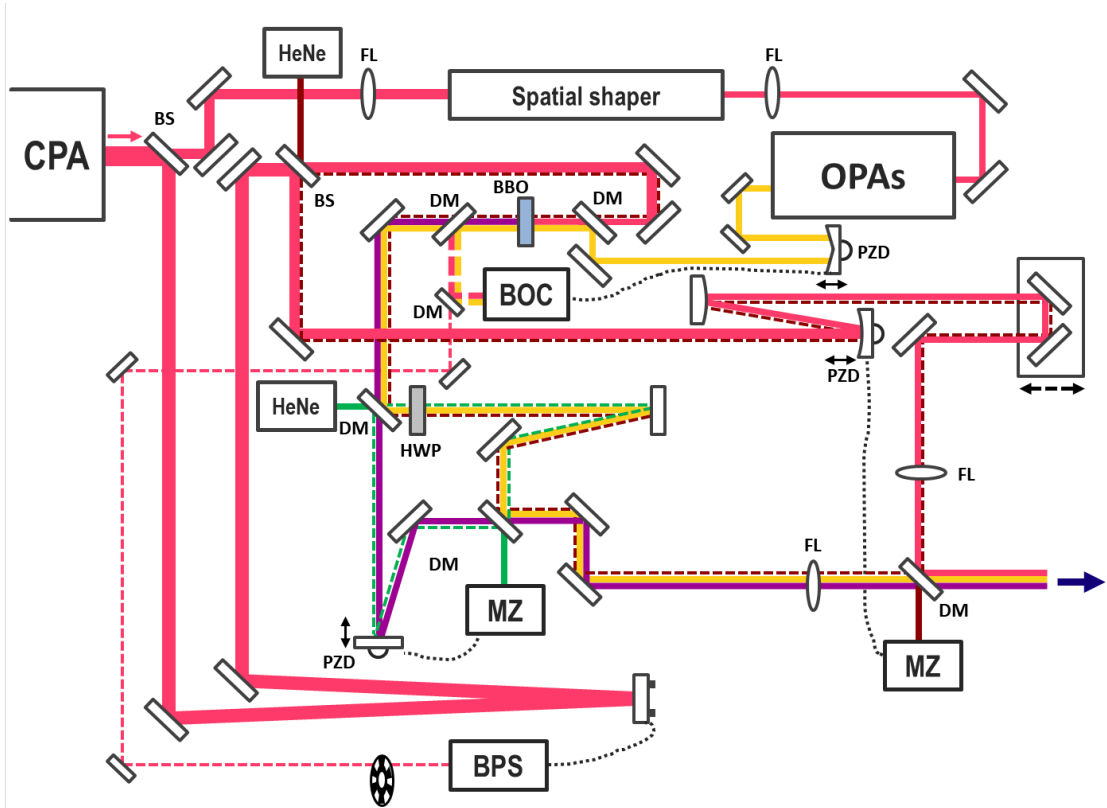

*Figure 1. Schematic of the three-channel waveform synthesizer.* CPA, chirped-pulse amplifier; OPA, white-light-seeded two-stage OPA; HeNe, He–Ne laser; BS, beam splitter; DM, dichroic mirror; BBO, beta barium borate crystal; FL, focus lens; HWP, half-wave plate; MZ, the signal detector of the Mach–Zehnder interferometer; BOC, balanced optical cross-correlator; BPS, beam-pointing stabilizer; PZD, piezo-transducer-actuated delay stage. Dotted lines indicate active feedback stabilization.

**Carrier-envelope phase stabilization**

To utilize higher pulse energy for the optical synthesizer, we employ a low-repetition rate (10-Hz) CPA laser. Even though CEP stabilization is commercially available for oscillators and kHz amplifier laser systems, CEP stabilization for a 10-Hz low-repetition rate laser is difficult because of the low-frequency rate of the active feedback signal generally. To realize CEP stabilization for a 10-Hz laser, we employed a laser system consisting of a CEP-stabilized oscillator, a 1-kHz front-end preamplifier, and a 10-Hz back-end power amplifier. The basic concept of our CEP stabilization method [30] is to sample the CEP reference of unamplified pulses from amplifier output, using a precisely synchronized optical chopper. Passing through the 10-Hz back-end power amplifier, the arrangement of each laser pulse in the temporal domain becomes a hybrid pulse train of repeatable 10-Hz pulses (amplified) and 990 shots/s (unamplified). After the pulse compressor, a portion of the hybrid pulse train is picked by a beam splitter. The chopper blocks the amplified high-energy 10-Hz laser shots and half the number of the 1-kHz seed pulse shots, then the passed-through 500-Hz unamplified seed pulses provide the same CEP information as the low-repetition CPA laser. By feeding back this 500-Hz CEP information to the compressor in the kHz amplifier, the low-repetition high-power laser is actively and indirectly CEP stabilized. Using this strategy, we can improve the active feedback bandwidth for CEP even though the main laser pulse with a high energy is operating at 10 Hz. The stabilized laser CEP has a low shot-to-shot noise of 453 mrad RMS, as shown in *Figure 2*.

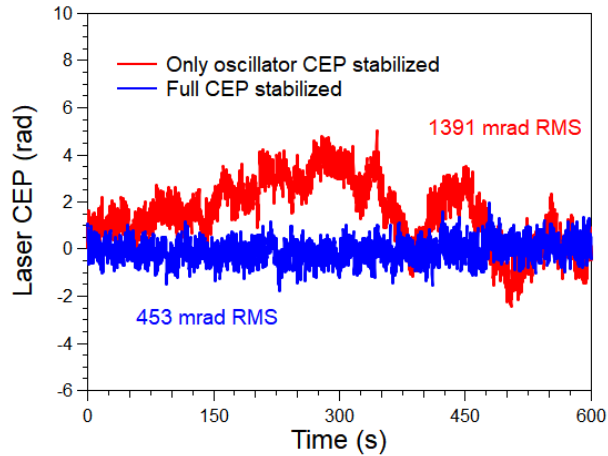

- *Figure 2. Laser-CEP stabilization.* *10-Hz single-shot CEP performance from the laser source measured using an f-to-2f interferometer; the red curve shows the 10-Hz CEP performance when CEP is stabilized only in the oscillator; the blue curve shows the full CEP performance when both oscillator CEP and amplifier CEP are stabilized.*

**Mach–Zehnder interferometer stabilization**

Optical delay variations due to mechanical vibrations among all the delay lines usually increase fast when the optical path is longer. To stabilize this type of delay jitter, two helium–neon continuous-wave (CW) laser beams are directly introduced into the synthesizer optical path from the backside of the beam splitter or dichroic mirror (shown in *Figure 1*). The synthesizer scheme in *Figure 1* contains two MZ interferometers, each of which can be actively stabilized by detecting the interference pattern of a CW laser beam transmitted through the MZ interferometer. To stabilize the delay jitter between pump and signal pulses, a He–Ne laser with a wavelength of 633 nm is injected at the pump-beam splitter. After passing parallelly through the pump and signal-beam paths, the beams are recombined at the final dichroic mirror. Then, a delay-dependent fringe pattern can be observed after the recombination. After a spatial linear slit filter, the fringes' movement transfers to intensity fluctuations depending on the delay changes, which can be monitored with a photodiode. We introduced an active feedback circuit utilizing the information of the recorded intensity fluctuations as an error signal. By closing the feedback control loop to the piezo-controlled delay stage, which is installed in the pump path, a less than 20 as RMS suppressed delay jitter is obtained. Similarly, another He–Ne laser with a wavelength of 543 nm is injected at the first dichroic mirror, where the signal pulses are separated from the idler for polarization change, then recombined again at the second dichroic mirror. With a similar feedback scheme to the idler-beam path, the delay jitter between signal and idler is suppressed. Detailed measured stabilization results are shown in *Figure 3(a)*. These jitter values are confirmed to be sufficiently small in stabilizing the synthesized waveform in our experiment for stable IAP generation.

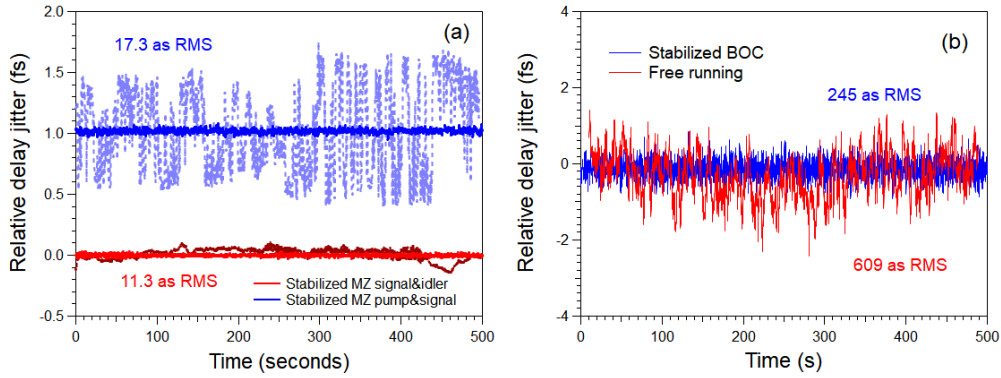

*Figure 3. Active stabilization of the delay jitters by the MZ and BOC. (a) Delay jitter stabilized using a Mach–Zehnder interferometer improved to 17.3 as (pump & signal), 11.3 as (signal & idler) RMS; (b) the delay jitter stabilized by BOC improved from 609 as to 245 as RMS.*

**Balanced optical cross-correlators**

Energy fluctuations of the laser pulse for OPA also introduce another type of delay jitter during the white-light generation process, which cannot be compensated by the MZ interferometer. The energy fluctuations of our 10-Hz CPA laser are measured to be around 2% RMS, which corresponds to a delay jitter of several femtoseconds. In this work, the white-light generation only appears in the first stage of OPA, which introduces delay jitter between pump and signal pulses. On the other hand, the delay jitter between signal and idler pulses is considered to be zero in principle excluding the delay jitter induced by the mechanical vibrations on its delay line. To compensate directly for the delay jitter introduced in the white-light generation process in the OPA, we evaluate it using the BOC technique, which is shown in *Figure 1*. The details of the BOC setup are similar to the one described in reference [17]. Experimentally, the delay jitter between pump and idler pulses is suppressed to 245 as RMS by the BOC setup, shown in *Figure 3(b)*. By monitoring the BOC signal between the signal and idler pulses while the MZ interferometer feedback is active, the delay jitter is measured to be 137 as RMS. This value is larger than the MZ interferometer stabilization due to the larger intrinsic noise of the BOC measurement because the BOC signal is collected from 10-Hz low-repetition rate laser pulses with sum-frequency generation.

**Beam-pointing stabilizer**

Even though the output beam pointing is stabilized by the stabilizer after the amplified laser source, a slight beam-pointing drift of the beams focused in the gas cell for HHG after transmitting through the synthesizer system was observed during long-term operation. Even more, the beam-pointing drift direction is observed to be different for the pump and idler beams. Therefore, beam overlapping will deteriorate this issue during long-term operation, which will cause the stability of the synthesized optical waveform to get worse. We further introduced another pointing stabilizer to control the pump-beam pointing at the focusing point by mirroring the beam path. Similar to CEP stabilization, the low-repetition rate beam-pointing stabilization is also a difficult issue due to the low feedback signal rate. We brought the same principle from the CEP stabilization, introducing an optical chopper to block the intense 10-Hz amplified pulse, while only letting the unamplified 500-Hz pump pulse pass through. With the beam-pointing information of the 500-Hz unamplified pump pulse, the beam pointing of the 10-Hz pump pulse is successfully stabilized. On the other hand, the signal-beam pointing is relatively stable because the signal beam is generated through OPA, the pumping pulse for the white-light generation and first two OPA stages is focused on a pinhole for spatial filtering in advance. Then, the pumping pulse on this pinhole acts as a new and stable beam-pointing source instead of the laser source; later, the focusing point of the signal beam for HHG will be an image point of the spatial filtering point by the focus lens. Thus, the signal-beam pointing is quite stable. The idler pulse is generated through the third stage of OPA, the beam pointing of the idler pulse is also passively drifting in the opposite direction of the pump pointing drift due to the phase matching. Therefore, by the pump-beam stabilization, the idler-beam pointing can also be indirectly stabilized.

- *Table 1. Stabilized parameters for the three-channel waveform synthesizer.*

| RMS | *Pump* | *Signal* | *Idler* |
|---|---|---|---|
| *Delay jitter (MZ)* | 17.3 as | 11.3 as ||
| *Delay jitter (BOC)* | 245 as | 137 as ||
| *Phase jitter* | 816 mrad | 106 mrad ||
| *Energy* | 1.48% | 2.1% | 2.66% |
| *CEP* | 453 mrad | 572 mrad | constant |

3. Results

Because the fully stabilized three-channel synthesizer is successfully constructed, we employ the loose-focusing geometry (the synthesizer output pulses are focused by two separate focus lenses: 4.5 m for the pump, 3.5 m for both signal and idler pulses) with a long medium (the interaction length: 8 cm, Ar pressure: 2.8 Torr) for scaling up the output harmonic energy [32]. ***Figure 4(a)*** shows the generated harmonic spectra in single-shot when one- and three-color pulses are used. With one-color (pump-pulse) input, the output spectrum shows clear discrete harmonics with the peaks' intensities gradually reducing until the end (35th order, 54 eV) in the cutoff region. In contrast, with the three-color input, the comb-like pattern in the spectrum disappears, a further extended smooth supercontinuum spectrum (50–70 eV) in the cutoff region is obtained. This supercontinuum spectrum supports an IAP with approximately 170 as transform-limited pulse duration. In this setup, the energies of the pump, signal, and idler pulses are optimized at 20.3, 4.3, and 1.6 mJ, respectively. Through the three-color waveform synthesis, the enhancement of the cutoff spectrum intensity reaches one to two orders of magnitude. Thanks to the very low gas pressure and low ionization probability (below 1%), the phase matching could be well achieved. With the $\approx 1 \times 10^{14}$ W/cm$^2$ focused intensity, plasma formation is avoided by staying in a regime with low ionization probability. The low gas pressure and low ionization also contribute in maintaining the optimized optical waveform during the HHG process. By comparison with our previous experimental results [32], the measured continuum soft-x-ray energy is evaluated to be at least 240 nJ. This indicates that the generated IAP reaches GW-level with 170 as duration. Under fully optimized phase-matching conditions with neutral media, we can expect the conversion efficiency to attain the $10^{-5}$ level. When the input synthesized waveform is stabilized, we can obtain a stable high-order harmonics continuum. The histogram of the integrated intensity for the single-shot spectrum around the cutoff region (50–70 eV) is shown in ***Figure 4(b)***. The RMS value is measured to be 4.1% when all the stabilizers are active (CEP, MZ, BOC, beam-pointing stabilizer), while the intensity stability declines to 15.1% without the waveform stabilization. In *Figure 5*, a broader cutoff spectrum at higher photon energy is generated by changing the gas medium from argon to neon (5 cm, 8.8 Torr) with the synthesizer input (25, 4.3, 1.6 mJ for pump, signal, idler, respectively). The Fourier-transform-limited pulse duration from this cutoff spectrum (100–130 eV) is 106 as. Taking into account our previous experiment [33], the evaluated soft-x-ray energy is about few-tens nJ. The lower energy-conversion efficiency is due to the neon atom's higher ionization energy and smaller recollisional cross section; therefore, the population of the tunnel-ionized electrons is reduced [34-36].

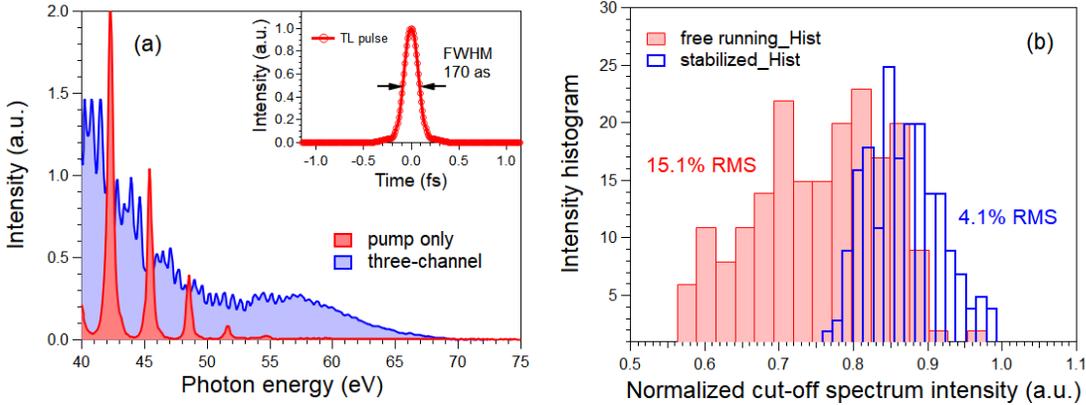

- *Figure 4. Generated spectrum of high-order harmonics supercontinuum from Ar. (a) Single-shot output spectra measured using one- and three-channel input; the inset shows the Fourier-transform-limited pulse from the cutoff continuum spectrum (50–70 eV); (b) The histogram profiles of the integrated intensity over the whole cutoff spectrum (50–70 eV) with/without stabilization of the synthesizer.*

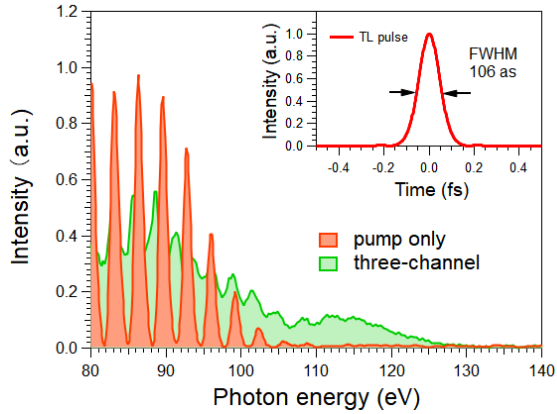

*Figure 5. Generated spectrum of high-order harmonics supercontinuum from Ne. Single-shot output spectra measured using one- and three-channel input; the inset shows the Fourier-transform-limited pulse from the cutoff continuum spectrum (100–130 eV).*

At this point, the cutoff energy enhancement effect of HHG enabled by the three-channel synthesizer is obvious, the theoretical expectation [20,21] is proved by experiment. Though there is no direct way to judge the waveform stability of the synthesizer, the generated HHG continuum spectrum allows us to evaluate it. Therefore, we designed a scanning scheme to prove it. In *Figure 6*, we change the delay of the pump or idler pulse using piezo-actuated delay stages, while maintaining other parameters of the stabilized synthesizer. From the experimental results, we observed the cutoff spectrum intensity to be periodically enhanced, depending on the delay change of the pump or idler pulse. This period is 1.3 and 3.4 fs corresponding to pump and idler, respectively. The full stabilization could not be activated while scanning the pump/idler-pulse delay, thus an intensity-fluctuating, unsmooth delay map is observed in the HHG spectra in *Figure 6(a)* and *(b)*. The delay-time dependence gives strong experimental proof of the success of intense optical-waveform stabilization.

In particular, see *Equation (1)*, the synthesized optical waveform should show a periodically repeating profile when one of the component multicycle pulses' delay changes. This repeatable period should be expected to be one cycle (2.6 fs for the pump, 6.8 fs for the idler) because the component pulse is a cosine-function-like waveform. So we need to understand the observed half-cycle periodical delay dependence during the HHG process. We start by considering it from the synthesized waveform based on its particularity, i.e., the fact that it is a *multicycle* pulse optical-waveform synthesizer. When the delay change ($\delta t_m$) is much smaller than the pulse duration ($\tau_m$), the synthesized electric field in *Equation (1)* can be rewritten as:

$$E_{3c} = \sum_{m=p,s,i} E_m \exp\left[-2\ln 2 \left(\frac{t}{\tau_m}\right)^2\right] \cos(\omega_m t + \omega_m \delta t_m + \varphi_m). \tag{2}$$

In this case, changing the delay of a component pulse (pump, signal, or idler pulse) within a small range does not affect the envelope of the waveform, but only affects its phase term ($\omega_m \delta t_m + \varphi_m$) inside the cosine function. This indicates that, when all other parameters are well stabilized in the synthesizer and the initial phase ($\varphi_m$) has a fixed value, then a delay change ($\delta t_m$) has an equivalent effect to a phase term variation ($\omega_m \delta t_m$) for each component pulse. This feature is meaningful in a real experimental operation. Usually, it is not easy to know the absolute phase of an electric field in a laser system. Only relative phase information can be measured with a designed detector, such as an f-to-2f interferometer [34]. This half-cycle periodic feature can help us precisely to control the synthesized optical waveform by changing the phase term of each component pulse through its delay change in a short range. This means an electric field close to the "perfect driver waveform" can be easily optimized by adjusting the delays of the component pulses. With a stabilized waveform output, an enhanced and stable supercontinuum spectrum can be obtained through HHG, which also indicates a stable intense IAP generation.

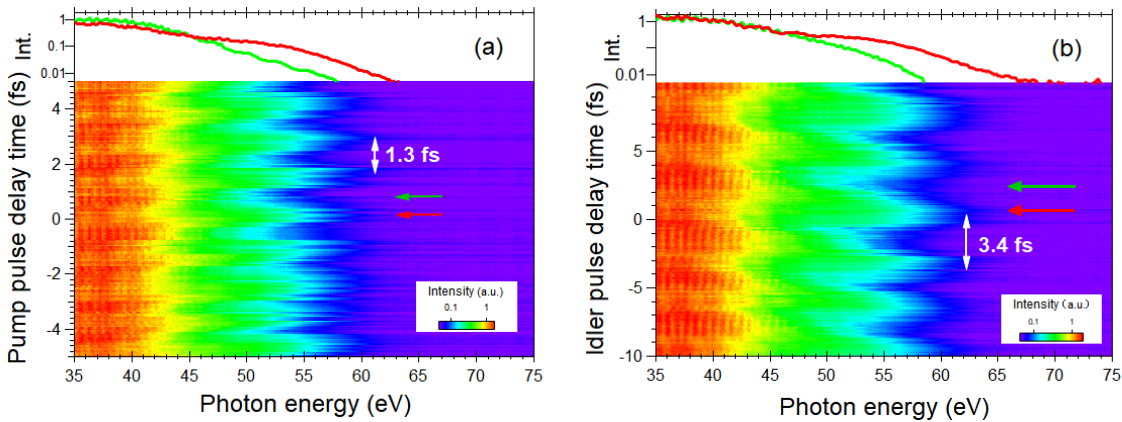

- *Figure 6. Experimental delay dependence of single-shot Ar HHG spectrum. (a) only changes the pump-pulse delay; (b) only changes the idler-pulse delay. The upper insets show the intensity profiles of the HHG spectrum marked in the delay maps.*

Reconstructing the synthesized waveform from experimental facts may help our understanding. The experimentally measured parameters of each pulse (wavelength, intensity, duration, etc.) are substituted into *Equation (2)*, then a synthesized waveform can be obtained as shown in *Figure 7(a)*. Here, the CEPs of the component pulses are first assumed to be zero. All the parameters except the absolute values of CEPs for each component pulse can be determined because the measured phases are relative. Therefore, back-tracking the absolute values of the CEPs should be done first. A hint comes from the cutoff spectrum, which has the relationship with the pumping source given by [35-37]: $E_{\text{cut-off}} = I_p + 3.17\, U_p$. Here, $E_{\text{cut-off}}$ is the maximum photon energy of the cutoff spectrum, $I_p$ is the ionization potential of the medium atom, $U_p$ is the electron quiver energy, which is related to the driving electric field. This formula is explained based on the semiclassical three-step model [35-37]. Note that the maximum intensity here is an instantaneous value, not the common meaning of the textbook-defined cycle-averaged quantity. For example, in *Figure 7(b)*, the related instantaneous intensity profiles $E^2(t)$ of the synthesized waveform and pump-pulse waveform are calculated from their electric-field waveforms $E(t)$ in *Figure 7(a)*. According to the three-step model, the HHG process starts to occur only when the driving field intensity is above the medium's ionization potential. The synthesized waveform produces a central peak with a maximum intensity that lasts only 0.32 cycles. Compared with the pump-pulse electric field, the intensity ratio is increased from 1.02 to 2.1 between the maximum peak and its nearest-neighbor peaks. This ratio needs a 2.5-fs pulse duration equivalently when only an 800-nm driving pulse is used. Therefore, the synthesized waveform generates an isolated pulse, in contrast, a pulse train is generated when only a pump pulse is used as the driving field. This has been experimentally confirmed in the generated continuum spectrum where the comb-like structure disappeared.

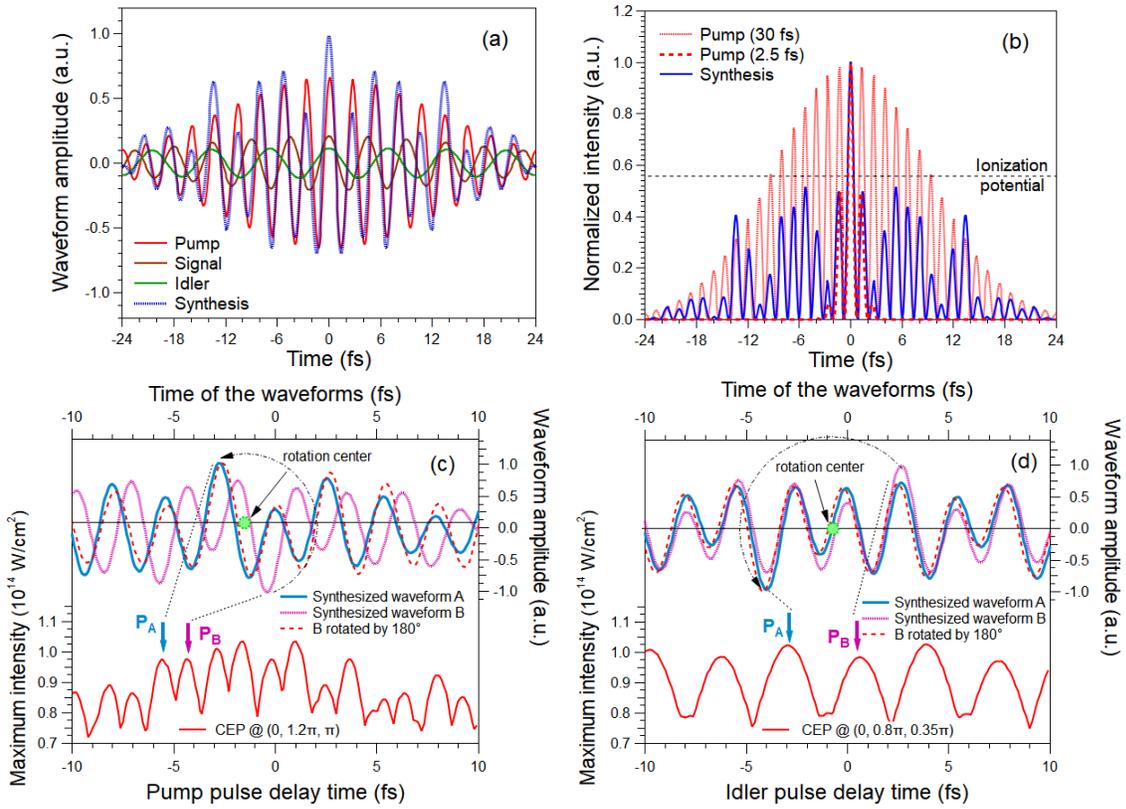

- *Figure 7. Simulated electric-field waveform with experimental parameters substituted. (a) Electric waveforms for the pump, signal, idler, and three-channel synthesis with (0, 0, 0) CEP combination; (b) corresponding instantaneous intensity profile $E^2(t)$ for pump and synthesized waveform in (a); for reference, the profile of a pump pulse with 2.5 fs duration is shown as the red dotted curve; (c, d) the lower red solid curve shows the maximum intensities backtracked from the synthesized waveforms by changing the pump/idler-pulse delay, upper solid curves correspond to the synthesized waveforms when the pump-pulse delay is at –5.6 and –4.3 fs, and idler-pulse delay is at –2.9 and 0.5 fs, marked with arrows of A and B; the red dotted curve is waveform B rotated by 180° around the indicated rotation center, which overlaps with the synthesized waveform A quite well around the maximum peaks ($P_A$ and $P_B$).*

Based on the above conclusion of maximum photon energy correlated with the maximum intensity, numerical simulations can be applied to back trace the synthesized optical waveform. After exhaustive searching of all the CEP (pump, signal, idler) combinations within a $2\pi$ range, two sets were found that can perfectly reconstruct the half-cycle periodic dependence on the maximum intensities of synthesized waveforms. They are (0, $1.2\pi$, $\pi$) and (0, $0.8\pi$, $0.35\pi$). Here, the pump-pulse CEP is arbitrarily defined as 0 for simplicity, values of the signal and idler are extracted from all the various combinations by checking the half-cycle periodic property. As shown with the lower red solid curve in *Figure 7(c)*, when the initial CEPs are (0, $1.2\pi$, $\pi$), by changing the pump-pulse delay, the maximum intensities of simulated synthesized waveforms approximately agree with the half-cycle periodic property in *Figure 6*. Similarly, when the idler-pulse delay changes with the initial CEP (0, $0.8\pi$, $0.35\pi$) in *Figure 7(d)*, the maximum intensities show the half-cycle periodic property. Moreover, these are the only two possible sets by traversing all phases from 0 to $2\pi$. When the pump CEP is changed from zero, the relative CEPs of signal and idler pulse can be found similarly with some phase shift.

By now, we can easily understand the half-cycle periodical feature through the simulated synthesized waveforms. In the upper side of *Figure 7(c)* and *(d)*, the synthesized waveforms (named with A and B) are extracted from the arrow-marked delay points ($P_A$ and $P_B$) in the lower maximum intensity traces. For example, the pump delay at –5.6 and –4.3 fs, idler delay at –2.9 and 0.5 fs, respectively. We find that the extracted waveform B rotated by 180° can overlap waveform A quite well around the maximum peak, while the delay change between waveforms A and B is a half-cycle time of the pump/idler pulse. Obviously, the symmetric waveforms have the same maximum intensity, as mentioned before, the maximum photon energy of the cutoff spectrum is positively correlated with the maximum intensity of the driving electric field, therefore, after the HHG process, the symmetric waveforms should yield similar output spectra. All in all, under the given initial CEPs, by changing the delay of the pump/idler pulse after a half-cycle time, the symmetric waveform will be synthesized, which has the same output after the HHG process. Therefore, the final time-delay dependence on the HHG spectrum shows a half-cycle periodicity instead of a one-cycle periodicity.

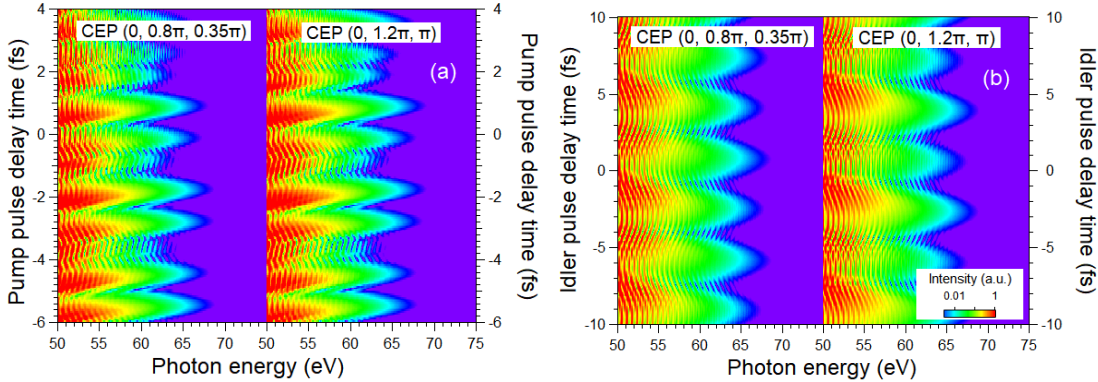

- *Figure 8. HHG spectrum profiles obtained with a theoretical calculation.* The HHG spectrum depending on the delay changes of the pump (a) and idler (b) for synthesized electric-field waveforms; (0, 1.2π, π) and (0, 0.8π, 0.35π) are both used as the absolute values of CEPs for each case.

Theoretical calculations are further applied to confirm the above analysis. The maximum kinetic energy of the cutoff continuum spectrum with pump/idler-pulse-delay change is calculated, as shown in *Figure S2*. The shapes of the curves agree with the maximum intensity distributions in *Figure 7(c)* and *(d)* very well. This calculation is based on the classical three-step model of the HHG process. The calculated HHG spectrum is shown in *Figure 8*, the results of delay-time dependences for the pump/idler pulse are in good agreement with the experimental results. The HHG spectrum calculations are based on the strong-field approximation model [38], and the collective response is calculated by solving the propagation equation [39,40]. The single-atom response induced by the three-color synthesized electric field, and the collective response of the macroscopic gas target to the laser driving field and high-harmonic fields [39,40] are considered. All the parameters are directly introduced from the experimental measurements, especially the CEP information is introduced as (0, 1.2π, π), and (0, 0.8π, 0.35π), which are deduced from the above analysis.

4. Discussion

In the previous section, the best CEP combinations for the most intense cutoff spectrum are obtained through back-tracking. Notably, the (0, 0, 0) CEP combination is not the condition of the cutoff continuum with the most intense and broadband spectrum, while it has the largest main-to-side-peak intensity ratio. To answer this, observation needs to connect it with the "perfect waveform" for generating the intense high-order harmonics. In the introduction, we mentioned that theoretical research [21,22] suggests that there is a most efficient waveform for driving the HHG. However, the "perfect waveform" from the theoretical perspective can also not be easily achieved in a real experiment. In this work, the resolution of the synthesized waveform on a time scale is limited by the small number of channels of different frequencies for synthesis. We can just obtain a "better waveform" with a low resolution that is close to the "perfect waveform," through the result analysis by successively trying different synthesizer parameters.

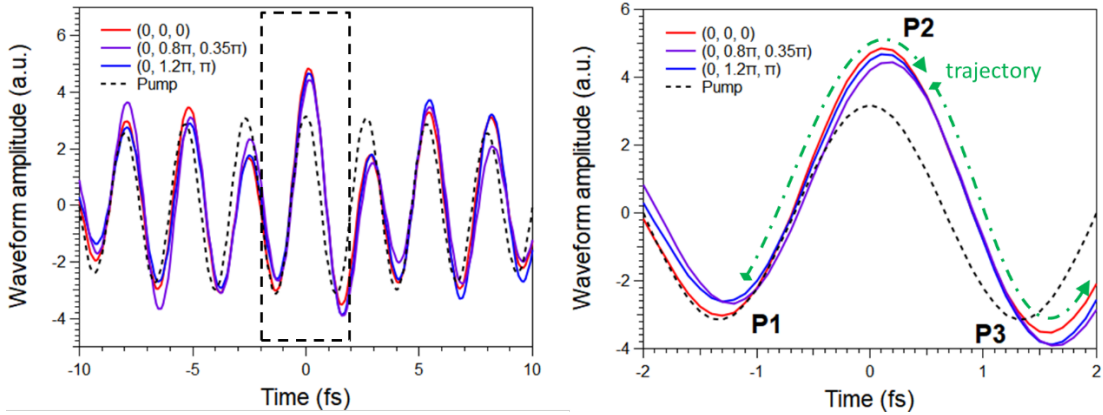

- *Figure 9. Synthesized electric-field waveform in amplitude with different CEP combinations.* The right panel shows the enlarged part from −2 fs to 2 fs time range in the left panel.

In *Figure 9*, we show the numerically reconstructed synthesized waveform under the CEP combinations of (0, 0, 0), (0, 1.2π, π), and (0, 0.8π, 0.35π) with the experimentally measured parameters; the pump-pulse waveform is also plotted as a dashed line for reference. We focus on the center part where the three most intense peaks of the waveform amplitude are located at around −1.3, 0.2, and 1.7 fs, where the electron could be ionized by the electric field, named as ionization points P1, P2, and P3, respectively. Each ionization point could be the starting point of one trajectory. According to the three-step model [35-37], for the P1 trajectory, the electron is ionized at P1 and returns to the parent ion about half a cycle later, i.e., at 0.7 fs. The electron's kinetic energy is highest for the CEP combination (0, 0, 0) because of the higher amplitude of P2. Similarly, the electron can be ionized at P2 and then returns to the parent ion at about 2 fs. For this P2 trajectory, the kinetic energy is lowest for (0, 0, 0) because of the lower amplitude of P3. More importantly, we should notice that the amplitude of P1 is lower than that of P2. This means that the ionization probability of the P1 trajectory is much lower than that of the P2 trajectory. In this case, the effective harmonic cutoff spectrum is determined by the P2 trajectory. Therefore, the cutoff spectrum intensity under the CEP combination of (0, 0, 0) is lower than that of the (0, 1.2π, π), and (0, 0.8π, 0.35π) combinations.

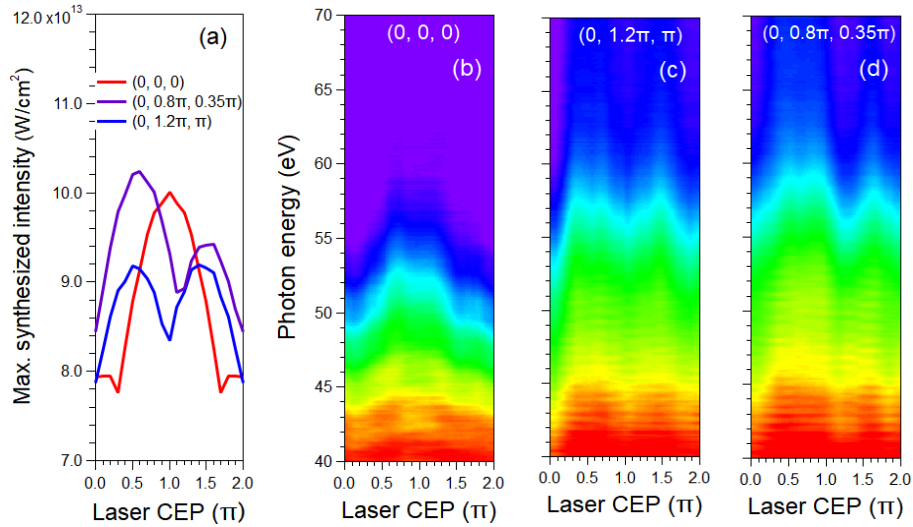

- *Figure 10. Laser-CEP dependence.* (a) Calculated maximum intensity profiles with laser-CEP changes, for the initial CEPs of (0, 0, 0), (0, 1.2π, π), (0, 0.8π, 0.35π); (b–d) measured single-shot HHG spectrograms corresponding to the three cases, respectively, using CEP tagging.

Usually, the driving-laser CEP dependence can be shown by a spectrum-scanning map when using a gating method or the two-channel synthesis for IAP generation. However, because of the characteristics of our parallel synthesizer, when the driving-laser CEP changes, the CEP of the idler pulse maintains a constant value due to the difference frequency generation process, while the CEPs of the pump and signal pulses change depending on the laser CEP. Thus, the performance of the driving-laser CEP dependence is varied. However, by assuming certain initial values of the CEPs, we can expect that the dependence on laser CEP is still interpretable. Similar to the previous simulation, the synthesized waveform maximum intensity depending on the laser-CEP change is shown in *Figure 10(a)*, for given initial CEP combinations of (0, 0, 0), (0, 1.2π, π), and (0, 0.8π, 0.35π). The maximum intensity predicts three different patterns when laser-CEP changes. During a 2π change, the values of the synthesized maximum intensity oscillates from minimum to maximum only once when the CEPs are (0, 0, 0), and it repeats twice when the CEPs are (0, 1.2π, π) and (0, 0.8π, 0.35π). Moreover, a slight difference in the intensity distribution is found between the (0, 1.2π, π) and (0, 0.8π, 0.35π) cases. According to the conclusion in the previous section, the extension of the generated continuum spectrum is related to the maximum intensity. We can expect the HHG spectrum to exhibit a similar feature on laser-CEP dependence. As shown in *Figure 10(b–d)*, measured using the CEP-tagging method, three different patterns of laser-CEP dependence are found experimentally, which are considered to correspond to the cases of initial CEPs (0, 0, 0), (0, 1.2π, π), and (0, 0.8π, 0.35π), respectively.

Another important fact that needs to be pointed out is that the value of the pump-laser CEP changes every day depending on where it is stabilized. This means the first absolute phase of the CEPs component usually is a nonzero value. Depending on the stabilization, the CEP of the signal and idler pulses may also vary. In our optical synthesizer, by tuning the delay of the pump/idler pulses, we could finally find the most intense output on the HHG spectrum. Through this procedure, the synthesized waveform will always be adjusted toward the same waveform, which is close to the "perfect waveform." This tunable, repeatable feature is confirmed by the different-day experimental results.

Certainly, this feature is based on the success of the full stabilization of the synthesizer. We would like to highlight this feature because it can be realized only in a multicycle parallel waveform synthesizer. When a few-cycle synthesizer changes the delay of its component pulse, because the component pulse duration is short, the approximation condition from *Equation (1)* to *Equation (2)* is not satisfied anymore, the synthesized waveform intensity envelope will be quickly changed depending on the delay.

5. **Conclusion**

Output performance and controllability of a TW-class parallel three-channel waveform synthesizer are reviewed, including multiple stabilization methods such as low-repetition-rate laser-CEP stabilization, MZ interferometer, BOC, and beam-pointing stabilizer. A fully stabilized intense synthesized optical waveform is successfully achieved. Through the HHG process employing a loose-focusing geometry, an intense shot-to-shot stable supercontinuum in the soft-x-ray region is generated. The continuum spectrum (50–70 eV) generated in an argon gas cell supports an isolated attosecond pulse with a TL duration of 170 as, energy above 240 nJ, which allows the generation of a GW-scale IAP. A shorter IAP in a higher photon energy range (100–130 eV) with 106 as TL duration, few-tens nJ, is also generated in a neon gas cell by the three-channel synthesizer. Through this study, a huge cutoff enhancement of the HHG spectrum using waveform synthesis is obtained experimentally and the underlying physics is unraveled theoretically. The multicycle pulse is proved not only to be capable of generating IAPs using waveform synthesis even though the pulse duration is around 10 times longer than that of few- or subcycle pulses but also to have the unique ability for energy scaling up to the GW class. Currently, another potential method to obtain a very high-flux IAP is considered to be the x-ray free-electron laser (XFEL) [41-43]. However, the accessibility of XFEL facilities is very limited. Laser-based intense IAP sources based on high-power waveform synthesizers are irreplaceable for providing a complementary nature with XFEL sources [44].

In addition, synthesis of multicycle pulses exhibits a new ability to tune the phases of each component pulse freely through changing its delay. Multicycle and multichannel synthesis may bring more degrees of freedom on experimentally achieving a "designed" electric field (such as the "perfect waveform" for IAP generation) through lasers, which may be used for other high-intensity electric-field research, such as laser electron accelerators [45,46] and strong-field physics [20,47]. The complexity of the synthesizer system will be increased dramatically when more channels are introduced because the number of parameters that need to be stabilized is increased. When the system complexity increases, program-controlled stabilization methods in modular form might be developed. The demand for the precision for the synthesized waveforms will push progress in this direction.

**Acknowledgments**

**Author contributions:**
"B.X., Y.T., and E.J.T designed the experimental setup and performed the experiments."
"Y.F. contributed to the development of the CEP-stabilized Ti:sapphire laser system."
"H.Y. and P.F.L. provided the theoretical calculations of HHG."
"B.X. and E.J.T. discussed all the experimental results and wrote the manuscript, which was polished by all authors."
"E.J.T. conceived the experimental idea and supervised this project as a whole."

**Funding:**

This work was supported, in part, by the Ministry of Education, Culture, Sports, Science and Technology of Japan (MEXT) through Grants-in-Aid under grants 17H01067, 19H05628, and 21H01850, in part by the FY 2019 Presidents Discretionary Funds of RIKEN, and in part by the Matsuo Foundation. B.X. acknowledges financial support from RIKEN for a Special Postdoctoral Researcher. Y.F. acknowledges support by the National Natural Science Foundation of China (92050107, 61690222); Major Science and Technology Infrastructure Preresearch Program of the CAS (J20-021-III); Key Deployment Research Program of XIOPM (S19-020-III). K.M. acknowledges support by the MEXT Quantum Leap Flagship Program (MEXT Q-LEAP) grant number JP-MXS0118068681. P.L. acknowledges support by the National Key Research and Development Program (2017YFE0116600) and the National Natural Science Foundation of China (91950202). O.D.M. acknowledges support by the priority program QUTIF (SPP1840 SOLSTICE) of Deutsche Forschungsgemeinschaft.

**Competing interests:**
The authors declare that they have no competing financial interests.

## Supplementary Materials

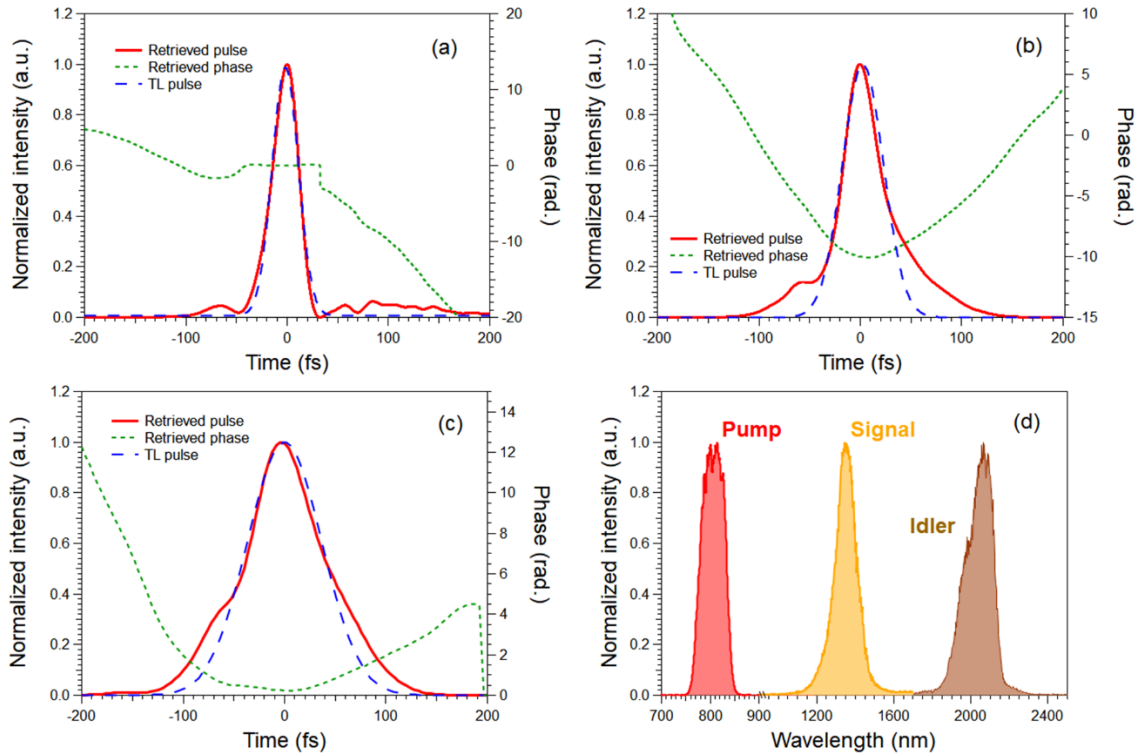

- *Figure S1. Component pulses composing the synthesized waveforms.* Pulse durations of the pump (a, 30 fs), signal (b, 44 fs), and idler (c, 88 fs) were experimentally measured using different methods. The retrieved temporal intensity is shown as the red solid curve, the retrieved temporal phase is plotted as the green solid curve. The calculated TL pulses are plotted as the blue dashed curve. The spectrum of the pump, signal, and idler component pulses are shown in panel (d).

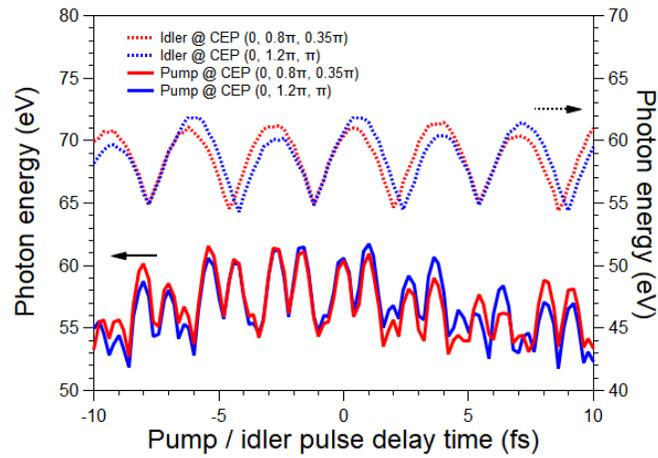

- *Figure S2. Calculated maximum kinetic energy profile of the HHG spectrum.* Solid lines show the calculated kinetic energy when pump-pulse delay is changed with different initial CEP combinations of (0, 1.2π, π) and (0, 0.8π, 0.35π); dotted lines show the kinetic energy when the delay of the idler pulse is changed.